\documentclass[useAMS,usenatbib]{mn2e}
\usepackage{natbib}
\usepackage{epsfig}
\usepackage{amsbsy}
\usepackage{pifont}
\usepackage{amsfonts}
\usepackage{hyperref}

\providecommand{\adsurl}[1]{\href{#1}{ADS}}

\newcommand{\lya}{Lyman-$\alpha$~}

\newcommand{\be}{\begin{equation}}
\newcommand{\ee}{\end{equation}}
\newcommand{\ba}{\begin{eqnarray}}
\newcommand{\ea}{\end{eqnarray}}
\newcommand{\brr}{\begin{array}}

\newcommand{\err}{\end{array}}
\newcommand{\bc}{\begin{center}}
\newcommand{\ec}{\end{center}}

\newcommand{\vel}{\,{\rm km\,s^{-1}}}

\newcommand{\mpch} {\rm $h^{-1}$ Mpc\,\,}
\newcommand{\ihmpc} {\rm $h$/Mpc\,\,}
\newcommand{\ns}{{\mbox{{\it n}$_{\rm s}$}}}

\newcommand{\teff}{{\mbox{$\tau_{\rm eff}$}}}

\newcommand{\msun}{{\rm M}_\odot}

\def\HI{\hbox{H$\,\rm \scriptstyle I\ $}}

\DeclareMathAlphabet{\mathsc}{OT1}{cmr}{m}{sc}
\def\testbx{bx}%
\DeclareRobustCommand{\ion}[2]{%
\relax\ifmmode
\ifx\testbx\f@series
{\mathbf{#1\,\mathsc{#2}}}\else
{\mathrm{#1\,\mathsc{#2}}}\fi
\else\textup{#1\,{\mdseries\textsc{#2}}}%
\fi}

\title[The impact of feedback on the \lya transmitted flux] {The impact of feedback from galaxy formation on the \lya transmitted flux}

\author[M. Viel, Joop Schaye \& C. M. Booth] 
{Matteo Viel$^{1,2}$, Joop Schaye$^{3}$ \& C. M. Booth$^{3,4,5}$ \\
$^1$ INAF - Osservatorio Astronomico di Trieste, Via G.B. Tiepolo 11,
I-34131 Trieste, Italy (viel@oats.inaf.it)\\
$^2$ INFN/National Institute for Nuclear Physics, Via Valerio 2, I-34127 Trieste, Italy\\
$^3$ Leiden Observatory, Leiden University, P.O. Box 9513, 2300 RA Leiden, the Netherlands\\
$^4$ Department of Astronomy and Astrophysics, The University of Chicago, Chicago, IL 60637, USA\\
$^5$ Kavli Institute for Cosmological Physics and Enrico Fermi Institute, The University of Chicago, Chicago, IL 60637, USA\\
}

\begin{document}
\maketitle
\begin{abstract}
The forest of \lya absorption lines seen in the spectra of distant
quasars has become an important probe of the distribution of matter in
the Universe. We use large, hydrodynamical simulations from the OWLS
project to investigate the effect of feedback from galaxy formation on
the probability distribution function and the power spectrum of the
\lya transmitted flux. While metal-line cooling is unimportant, both
galactic outflows from massive galaxies driven by active galactic
nuclei and winds from low-mass galaxies driven by supernovae have a
substantial impact on the flux statistics. At redshift $z=2.25$, the
effects on the flux statistics are of a similar magnitude as the
statistical uncertainties of published data sets.  The changes in the
flux statistics are not due to differences in the temperature-density
relation of the photo-ionised gas. Instead, they are caused by changes
in the density distribution and in the fraction of hot, collisionally
ionised gas. It may be possible to disentangle astrophysical and
cosmological effects by taking advantage of the fact that they induce
different redshift dependencies. In particular, the magnitude of the
feedback effects appears to decrease rapidly with increasing redshift.
Analyses of \lya forest data from surveys that are currently in
process, such as BOSS/SDSS-III and X-Shooter/VLT, must take galactic
winds into account.
\end{abstract}


\section{Introduction}
The \HI \lya forest, i.e.\ the multitude of \HI \lya absorption lines
seen in the spectra of distant quasars (QSOs), is an important
cosmological observable that probes matter density fluctuations in the
intergalactic medium (IGM) over a unique range of redshifts, scales
and environments. Many attempts have been made to measure physical
properties of the IGM using \lya forest data. The two most common
approaches are either based on decomposing the information encoded in
the transmitted flux via Voigt profile fitting or treating the flux as
a continuous field with directly measurable statistical properties
\citep[e.g.][]{rauch98,croft98,croft02,meiksin09}. In the second approach, measurement
of the zero-, one-, two- or three-point probability distribution
functions (i.e.\ the mean flux level, the flux probability
distribution function, the flux power and bispectrum) enable a variety
of physical properties to be explored. The mean flux level for
example, is sensitive to the amplitude of the meta-galactic Ultra Violet (UV)
background \citep[e.g.][]{rauch97,tytler04,bolt05} while the flux
probability distribution function (PDF) is sensitive to the thermal
evolution of the IGM \citep[e.g.][]{theuns00,mcquinn09,bolton09,peeples10,peeples10b}. The
flux power spectrum has been used to constrain cosmological parameters
and the behaviour of dark matter (DM) at small scales
\citep{croft99,vhs,seljak06,viel08}, the flux bispectrum can be used to search
for signatures of non-Gaussianities in the matter distribution
\citep{viel09} and wavelet techniques can also be applied in order to
constrain the IGM temperature
\citep[e.g.][]{meiksin00,theuns02b,garzilli12}.

The data used for these investigations consist mainly of two sets of
QSO spectra: the SDSS low-resolution, low signal-to-noise sample and
UVES/VLT or HIRES/KECK samples of high-resolution spectra.  The
characteristics of the low- and high-resolution data sets are very
different (the number of SDSS spectra is a factor $\sim 200$ larger
than that of high-resolution samples, but the latter probe smaller
scales due to the higher spectral resolution). Measurements based on
\lya forest data have reached a level of accuracy where an
understanding of systematic uncertainties at the percent level or
below (the magnitude of statistical errors associated with the SDSS
sample) has become important.

In this work we use two sets of simulations to investigate the effects
of various physical and cosmological processes on \lya flux
statistics.  Firstly we use the suite of large cosmological
hydrodynamical simulations that comprise the OverWhelmingly Large
Simulations project (OWLS; \citealt{schaye10}) to investigate the
effect of physical processes such as metal-line cooling and galactic
winds driven by supernovae (SNe) and active galactic nuclei (AGN) on
the flux statistics. \citet{altay11} and \citet{tepper12} have shown
that the OWLS models reproduce the observed \HI column density
distributions at $z=3$ and $z=0$, respectively. Secondly, we use sets
of simulations in which cosmological parameters including the neutrino
mass \citep{vhs10}, the thermal history of the IGM \citep{viel09PDF}
and the effect of WDM \citep{vielwdm} are varied.

Throughout, we focus on the effects of these physical processes on two
statistics that have been shown to be very constraining: the flux PDF
and the flux power spectrum.  These statistics have previously been
investigated by several authors either in isolation
\citep[e.g.][]{mcdonald00,jena,Becker:2006qj,lidz06,tkim,bolton08pdf}, or
jointly \citep[e.g.][]{theuns00,meiksin01,zaroubi06,desjacques07}.  We
will compare with the present error bars derived from the flux PDF of
high-resolution UVES QSO spectra \citep{tkim} and the flux power
measured from the SDSS data \citep{mcdonald06}. Note that ongoing
surveys such as BOSS/SDSS-III will strongly reduce these statistical
error bars.

Previous simulation studies have found the effect of feedback on the
\lya forest to be relatively small, because the winds tend to escape
into the voids, leaving the filamentary structures responsible for
most of the \lya forest nearly intact
\citep[e.g.][]{theuns02,bruscoli03,mcdonald2005winds,cen2005,kollmeier06,bertone06,borgani09,tornatore10,kawatarauch,tepper12}. However,
the precision expected from upcoming surveys implies that even small
differences may be important. Moreover, previous work has generally
considered only relatively weak feedback processes and as a result has
suffered from ``overcooling''.  Inclusion of the strong feedback that
appears to be required to reproduce observations may lead to much
stronger effects and qualitative differences, as was recently found
for the low-$z$ matter power spectrum \citep{vandaalen11,semboloni11}.
Indeed, the OWLS models have already been used to demonstrate that
feedback from galaxy formation changes the density profiles of haloes
\citep{duffy10} and the matter power spectrum \citep{vandaalen11} to a
degree that is highly relevant for upcoming weak lensing experiments
\citep{semboloni11}. Furthermore, by using a set of hydrodynamic
simulations, galactic feedback either in the form of AGN feedback
and/or winds has also been shown to affect the properties of Damped
Lyman-$\alpha$ systems and metal systems like CIV
(e.g. \cite{tescari09,tescari11,barai12}) and the properties of the
intracluster medium (e.g. \cite{fabjan}).

Recently, \citet{chang12} argued that plasma instabilities cause
ultra-relativisitic pairs produced by blazars to dissipate their
energy in the IGM and that this process dominates over photo-heating
at low densities. \citet{puchwein12} demonstrated that this heating
mechanism has the potential to significantly affect the observed \lya
forest flux statistics. Although we will confine ourselves to more
traditional, and more localised, feedback effects, it should be kept
in mind that other feedback mechanisms, such as blazar heating, may
also be important.

Here we demonstrate that the effects of outflows driven by feedback
effects from galaxy formation may already be comparable to the
statistical errors of past and current surveys. We find that the
effect of feedback decreases rapidly with increasing redshift, which
may make it possible to disentangle astrophysical and cosmological
effects.

The plan of the paper is as follows: in Section 2 we describe the
simulations used and the physics implemented, in Section 3 we compare
models that include different feedback processes and we compare the
magnitude of the feedback effects to that of changes in the
cosmological parameters and in the thermal history of the IGM.
Finally, in Section 4 we present a summary and discuss future
perspectives.

\section{Simulations}

\begin{table*}
\caption{Summary of the OWLS models used in the present analysis.  The
  simulations that we analyse differ from one another only in terms of
  the included feedback processes, and only these parameters are noted
  here.  Each simulation contains $512^3$ particles of both DM and
  gas. A more complete description of the physics included in each
  simulation is present in Sec.2.  From left-to-right, the columns
  contain: (1) simulation identifier, (2) Whether the simulation
  treats AGN feedback, (3) Whether the simulation includes cooling
  through metal lines, (4) Whether or not the simulation contains SN
  winds, and for the simulations that do contain SN winds, columns (5)
  and (6) list the wind mass-loading and wind velocity. The
    symbols $\rho$ and $v_{\rm c}$ indicate the gas density local to
    a feedback event and the host halo circular velocity, respectively.}
\begin{tabular}{llllll}
\hline
Simulation$^{(1)}$ &  AGN$^{(2)}$ & Z-cooling$^{(3)}$ & SN Winds$^{(4)}$ & mass-loading$^{(5)}$ & wind-velocity$^{(6)}$ \\ 
\hline 
\emph{REF} & $\times$ & \checkmark & \checkmark & 2 & 600
\\
\emph{AGN} & \checkmark & \checkmark & \checkmark & 2 & 600
\\
\emph{NOSN\_NOZCOOL} &  $\times$ & $\times$ & $\times$ & -- & --
\\ 
\emph{NOZCOOL} & $\times$ & $\times$ & \checkmark & 2 & 600
\\ 
\emph{WDENS} & $\times$ & \checkmark & \checkmark & $\propto\rho^{-1/3}$ & $\propto\rho^{1/6}$
\\ 
\emph{WML1V848} & $\times$ & \checkmark & \checkmark & 1 & 848
\\                           
\emph{WML4V424} & $\times$ & \checkmark & \checkmark & 4 & 424
\\                           
\emph{WML8V300} & $\times$ & \checkmark & \checkmark & 8 & 300
\\
\emph{WVCIRC} & $\times$ & \checkmark & \checkmark & {\bf $\propto v_{\rm c}$} & {\bf $\propto v_{\rm c}^{-1}$}
\\
\hline
\end{tabular}
\label{tab1sims}
\end{table*}

\begin{table*}
\caption{Summary of the simulations used to probe cosmological
  parameters and the effect of the thermal state of the IGM.  Each
  simulation is compared to a reference $\Lambda CDM$ simulation,
  which is identical apart from any differences described in the
  fourth column.  From left-to-right, the columns list: (1) simulation
  identifier, (2) Comoving box size in Mpc/$h$, (3) Number of baryonic
  particles, (4) Brief description of the difference between each
  simulation and its corresponding reference case.}
\begin{tabular}{llll}
\hline
Simulation$^{(4)}$ & Box Size (Mpc$/h$)$^{(2)}$ & N$_{{\rm gas}}^{(3)}$ & Description$^{(4)}$ \\ 
\hline 
$WDM$ & $25$&$512^3$& Warm dark matter 1keV vs. $\Lambda$CDM
\\ 
$\nu$ & $60$&$512^3$& Neutrino mass $\Sigma m_{\nu}=0.6$ eV + $\Lambda$CDM vs.  massless neutrinos + $\Lambda$CDM
\\
$\gamma$ & $20$&$256^3$& $T-\rho$ relation of the IGM has slope $\gamma=1.6$ vs. $\gamma=1.3$
\\ 
$n_{\rm s}$ & $20$ &$256^3$& Spectral index $n_{\rm s}=0.95$ vs. $n_{\rm s}=0.9$
\\
$\sigma_8$ & $20$ &$256^3$& $\sigma_8=0.85$ vs. $\sigma_8=0.75$
\\
\hline
\end{tabular}
\label{tab2sims}
\end{table*}

In this work we use cosmological smoothed particle hydrodynamics (SPH)
simulations of representative volumes of the Universe taken from the
OWLS project \citep{schaye10}.  Gravitational forces and the
hydrodynamic equations are solved with an extended version of the
parallel PMTree-SPH code {\sc gadget iii} \citep[last described in
][]{springel}.  Our simulations assume a $\Lambda$CDM cosmology with
parameters from the 3-yr Wilkinson Microwave Anisotropy Probe (WMAP)
results\footnote{These parameter values are largely consistent with
  the WMAP 7-year results \citep{komatsu11}, with the main discrepancy
  being that $\sigma_8$ is 8\%, or 2.3$\sigma$, lower in the WMAP 3
  data than allowed by the WMAP 7 data.}, $\Omega_{\rm m} = 0.238$,
$\Omega_\Lambda=0.762$, $\Omega_{\rm b} = 0.0418$, $h$ = 0.73,
$\sigma_8 = 0.74$ and $n_{\rm s} = 0.951$.

In addition to gravity and hydrodynamic forces, the simulations use
sub-grid modules to follow the galaxy formation process. The sub-grid
modules used in each simulation include the modelling of star
formation, SN feedback, the growth of supermassive black holes (BHs)
and AGN feedback, radiative cooling and chemodynamics, as described in
\citet{schaye08}, \citet{dallavecchia08}, \citet{booth09},
\citet{wiersma08} and \citet{wiersma09}, respectively.

Table~\ref{tab1sims} contains a list of the OWLS simulations used in
this study.  We begin here by describing our fiducial model
(\emph{REF}) before discussing the variations in the sub-grid physics
used in the other simulations.  In the \emph{REF} simulation,
radiative cooling and heating are calculated element-by-element by
explicitly following the 11 elements H, He, C, N, O, Ne, Mg, Si, S, Ca
and Fe in the presence of the Cosmic Microwave Background and the
\citet{haardtmadau} model for the UV/X-ray background radiation from
quasars and galaxies, as described in \citet{wiersma08},\citet{wiersma09}.

Star-forming gas is modelled by imposing a polytropic effective
equation of state for dense particles ($n_{\textrm{\scriptsize H}} >
10^{-1}$ cm$^{-3}$).  These particles feel a pressure $P \propto
\rho^{\gamma_{\textrm{\scriptsize eff}}}$, where
$\gamma_{\textrm{\scriptsize eff}}=4/3$ is a polytropic index and
$\rho$ is the proper mass density of the gas.  Star formation is
followed stochastically \citep{schaye08}, with a pressure dependent
star formation rate chosen to match the observed relation between star
formation surface density and gas surface density \citet{kennicutt98}.

The \emph{REF} simulation does not include a prescription for BH
growth and AGN feedback, but it does treat feedback from SNe.  SN
feedback is implemented by injecting $\sim$40\% of the energy released
by Type II SNe locally as kinetic energy. Each newly formed star
particle kicks on average $\eta=2$ of its neighbouring gas particles
into the wind, isotropically at a velocity of $v_{\rm w}=600$\,km/s.
These two parameter values were chosen such that the peak in the
global star formation history approximately matches the observations
\citep{schaye10}. \citet{altay11} demonstrated that the reference
model agrees with the observed $z=3$ \HI column density distribution
over the full 10 orders of magnitude in \HI for which it is measured.

The total energy injected into SN-driven winds is proportional to
$\eta v_{\rm w}^2$, but we have some freedom in choosing how the
energy is distributed between the mass-loading and wind velocity.  We
therefore analyse an additional three simulations in which the
initial wind velocity and mass-loading are scaled such that the total energy
output per unit stellar mass is the same as in \emph{REF}, but the distribution across
mass-loading and wind velocity is different.  These three simulations
are termed \emph{WML1V848} ($\eta=1$, $v_{\rm w}=848$\,km/s),
\emph{WML4V424} ($\eta=4$, $v_{\rm w}=424$\,km/s), and \emph{WML8V300}
($\eta=8$, $v_{\rm w}=300$\,km/s).

The \emph{WDENS} simulation
is the same as the \emph{REF} simulation but in place of using a
constant wind velocity ($v_{\rm w}$) and mass-loading ($\eta$), both
parameters depend on the local gas density such that
\be
v_{\rm w}=600\,{\rm km\,s}^{-1}\,\Big(\frac{n_{\rm H}}{0.1\,{\rm cm}^{-3}}\Big)^{1/6},
\ee
\be
\eta=2\Big(\frac{v_{\rm w}}{600\,{\rm km\,s}^{-1}}\Big)^{-2}.
\ee
This ensures that in the densest regions winds are launched with the
largest speeds. For gas on the star-forming equation of state, this
scaling gives a wind speed that is proportional to the effective local
sound speed.  Note that the \emph{REF} and \emph{WDENS} models inject the same
amount energy per unit stellar mass. The differences in results
therefore result solely from how energy is distributed between mass
loading and wind velocity.

In addition, we consider a simulation that where the wind velocity and mass-loading scalings match those predicted by models of radiation pressure driven winds \citep{murray05}.  Other simulations that employ these \lq momentum driven\rq\, wind scalings have been shown to provide a good match to OVI absorption line statistics \citep{oppenheimer06}.  In this simulation, \emph{WVCIRC}, the wind velocity and mass-loading are given by
\be
v_{\rm w}=(5/\sqrt{2})v_{\rm c}\,{\rm km/s},
\ee
and
\be
\eta=(150/\sqrt{2})v_{\rm c}^{-1},
\ee
where $v_{\rm c}=\sqrt{GM_{\rm vir}/R_{\rm vir}}$ is the host halo circular velocity.

We caution that comparisons of our results to the simulations of \citet{oppenheimer06, oppenheimer08, oppenheimer09} are problematic because these authors varied the parameters of their models between papers \citep[see][for a discussion of this point]{schaye10}, and they temporarily \lq decouple\rq\, wind particles from the hydrodynamical forces, allowing winds to escape from the interstellar medium, which has a large effect on galaxy properties (see Haas et al. in prep. for a discussion of this point).

The \emph{NOSN\_NOZCOOL} simulation is identical to the \emph{REF}
simulation except that it does not contain feedback
from SNe and that gas cooling rates assume zero metallicity.
\emph{NOZCOOL} is the same as \emph{REF} except that all cooling rates
are calculated assuming primordial abundances.  We can therefore
isolate the effects of SN feedback by comparing simulations
\emph{NOSN\_NOZCOOL} and \emph{NOZCOOL}.  

With the exception of \emph{NOSN\_NOZCOOL}, all of the OWLS
simulations analysed here contain a prescription for SN feedback, but
the specific choices for the SN mass-loading and wind-velocity affect
the range of halo masses in which SN feedback is effective.
Prescriptions with high mass-loadings are able to eject a large amount
of gas from objects with low star formation rates, we therefore expect
that these prescriptions (\emph{WML4V424} and \emph{WML8V300}) will be
able to efficiently remove gas from small galaxies at high redshift.
These models could potentially have a large impact on the high-$z$
\lya forest as it is precisely the outflows from these objects that
have a large enough volume filling factor to account for observations
of metal-line absorption and hence to impact the high-$z$ IGM
\citep[e.g.][]{booth12}.

Prescriptions with high mass-loadings (and correspondingly low wind
velocities) are unable to eject gas from high-mass galaxies because
the winds are pressure-confined by the ambient interstellar medium
\citep{dallavecchia08} and because low wind velocities result in large
thermal losses in the dense interstellar medium of massive galaxies
\citep{dallavecchia12}.  Models that use a large wind velocity
(\emph{WML1V848}) are therefore able to eject gas from galaxies of
higher mass.  Comparison of these different simulations can therefore
inform us how winds escaping from haloes of different masses affect
the statistics of the IGM.

Finally, the \emph{AGN} simulation is the same as \emph{REF} except
that it includes BH growth and AGN feedback.  These two processes are
implemented using the method of \citet{booth09} which is, in turn, a
modified version of the method pioneered by \citet{springelbh}. Seed
BHs are inserted into the simulation by regularly running a halo
finder and placing small BHs ($m_{\rm seed}=10^5\,\msun$) into every
halo with mass $>10^{10}\,\msun$ that does not yet contain a BH.
These seed BHs then grow both through merging with other BHs and by
Eddington-limited gas accretion. Accretion rates in low-density gas
($n_{\rm H}<10^{-1}\,$cm$^{-3}$) are equal to the Bondi-Hoyle rate.
In star-forming gas, this accretion rate is boosted by a factor
$(n_{\rm H}/10^{-1}\,{\rm cm}^{−3})^2$ to compensate for the lack of a
cold, interstellar gas phase and the finite resolution (see Booth \&
Schaye 2009 for a discussion). For a given accretion rate,
$\dot{m}_{\rm accr}$, the BH growth rate is $\dot{m}_{\rm BH} =
(1-\epsilon_{\rm r})\dot{m}_{\rm accr}$, where $\epsilon_{\rm r}=0.1$
is the radiative efficiency of the BH.

The amount of energy available for AGN feedback is then given by
$\dot{E}=\epsilon_{\rm f}\epsilon_{\rm r}\dot{m}_{\rm accr}c^2$, where
$c$ is the speed of light and $\epsilon_{\rm f}$ is a free parameter.
Feedback is implemented by allowing the BH to build up a reservoir of
accreted energy until it is capable of heating one of its neighbours
by a temperature $\Delta T=10^8$~K.  This energy is then injected
thermally into one of the BH's surrounding gas particles.  The
parameter $\epsilon_{\rm f}=0.15$ is set by requiring that the
\emph{AGN} simulations reproduce the global BH density at $z=0$.

Each of the physics variations was run in boxes of comoving size 25\,\mpch and
100\,\mpch and contained $512^3$ particles of both gas and DM.  This
corresponds to an initial gas particle mass of
$1.35\times10^6\,\msun/h$ ($8.7\times10^7\msun/h$) in the 25 (100)
\mpch boxes, respectively.  In Appendix~\ref{appendix1resolution} we
use a further set of simulations that use smaller box sizes and
degraded numerical resolutions to demonstrate that our conclusions are
robust with respect to numerical resolution and box size effects.

We note that for all the OWLS models analysed in this paper, the
``equation-of-state'' (EOS) of the photo-ionised IGM is basically
unchanged and results in the same evolution for the
parameters $T_0,\gamma$ that describe it.

In order to compare the effects induced by feedback on the transmitted
\lya flux with those produced by cosmological parameters or different
thermal histories, we also consider a second set of hydrodynamical
simulations.  These simulations were also performed with the {\sc
  gadget iii} code but they include neither metal enrichment nor AGN
feedback and they model star formation using a simplified star
formation criterion that converts all the gas particles with $T<10^5$
K and densities larger than one thousand times the mean density
into stars \citep{vhs}.  This set of simulations is summarised in
Table~\ref{tab2sims}.  In brief, we consider: i) a warm DM model (WDM)
with a 1~keV mass for the DM particle as presented in \cite{vielwdm}.
This simulation is compared to a corresponding $\Lambda$CDM simulation
(these two simulations were performed in a 25 comoving Mpc$/h$ box
with $2\times 512^3$ DM and gas particles); ii) two simulations that
differ only in their choice of spectral index ($n_{\rm s}=0.95$) and matter
power spectrum amplitude ($\sigma_8=0.75$), relative to the
corresponding reference simulation ($n_{\rm s}=1$, $\sigma_8=0.85$).  These
simulations were performed in boxes of 20 comoving Mpc$/h$, and
contain $256^3$ particles of both DM and gas;  iii) two 20 comoving
Mpc$/h$, $2\times 256^3$ simulations in which the thermal history of
the IGM has different power-law indices for the temperature-density
relation of the low-density photo-ionised gas, namely
$T=T_0(1+\delta)^{(\gamma-1)}$, with $\gamma=1.3,1.6$ as in
\cite{viel09PDF}; iv) a simulation of 60 comoving Mpc$/h$ with $512^3$
massive neutrinos of total mass $\Sigma m_{\nu}=0.6$ eV and $512^3$ DM
particles compared with a simulation in which the neutrinos are
massless \citep{vhs10}.

The mass per gas particle is
$1.5\times10^{6}\,\msun$ for the WDM simulation, $5.8\times10^{6}\,\msun$ for the neutrino
simulation, and $6.1\times 10^{6}\,\msun$ for all the other cases.  This
second set of simulations is not homogeneous, so we will only use it
to highlight the relative differences with respect to the matched
reference simulation that corresponds to each simulation
($\sigma_8=0.85$, $n_{\rm s}=0.95$, $\gamma=1.3$ and $\Sigma m_{\nu}=0$).
The resolution of each of these simulations is sufficient to obtain
converged results for the \lya PDF at $z<3$
\citep{bolton08pdf,viel09PDF,calura012}.

\section{Flux statistics}

\begin{figure*}
\begin{center}
\includegraphics[width=16cm]{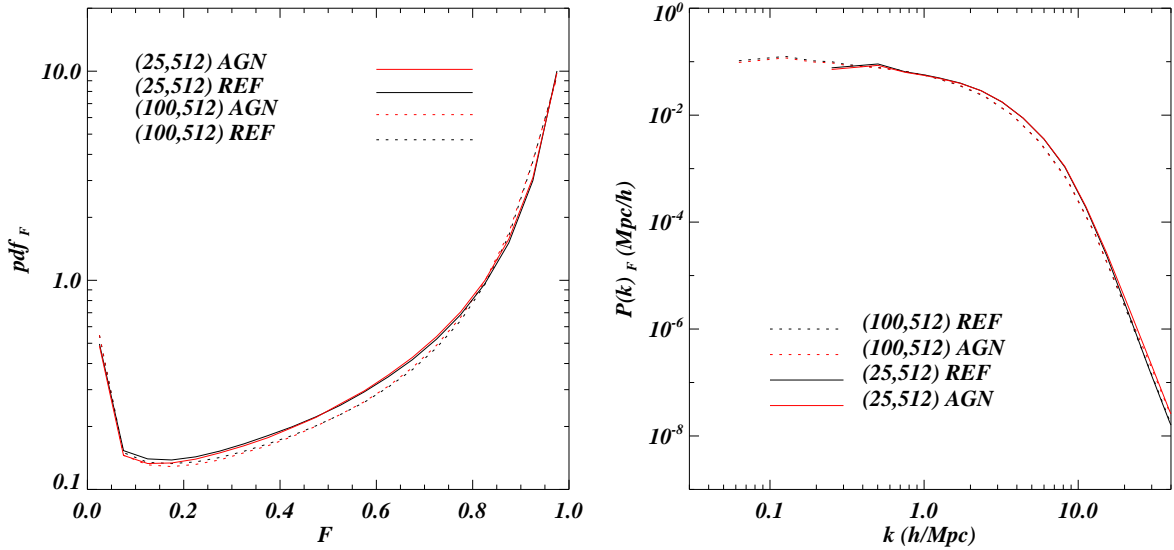}
\end{center}
\caption{Flux PDF (left) and flux power (right) for two different
  cosmological boxes (100,512) and (25,512) represented by the
  dotted and solid lines, respectively.  We plot the
  flux PDF and flux power for the \emph{AGN} (red curves) and
  \emph{REF} (black curves) simulations at $z=2.25$.}
\label{fig_a}
\end{figure*}

\begin{figure*}
\begin{center}
\includegraphics[width=16cm]{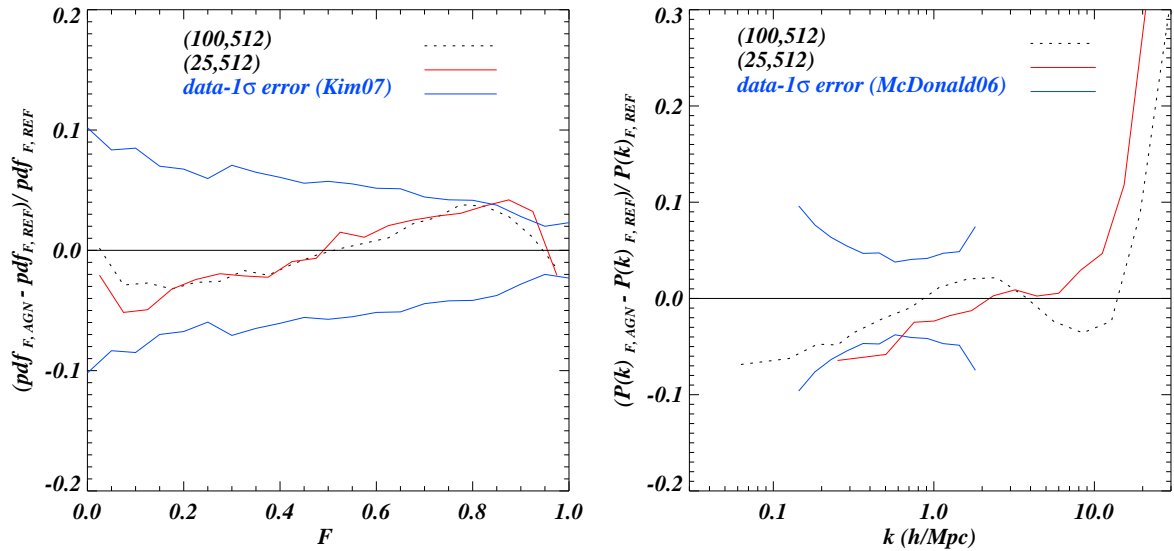}
\end{center}
\caption{Effect of \emph{AGN} feedback on the flux PDF (left) and
  flux power (right) for two different cosmological boxes (100,512)
  and (25,512) represented by the black, dotted and red, solid lines,
  respectively.  We plot the fractional differences between the flux
  PDF and flux power of the \emph{AGN} and \emph{REF}
  simulations at $z=2.25$. The blue, solid lines bracket the
  $1\sigma$ statistical observed errors from UVES/VLT at $z=2.07$ PDF
  \citep{tkim} and SDSS flux power at $z=2.2$ \citep{mcdonald06}.}
\label{fig_b}
\end{figure*}

\begin{figure*}
\begin{center}
\includegraphics[width=16cm]{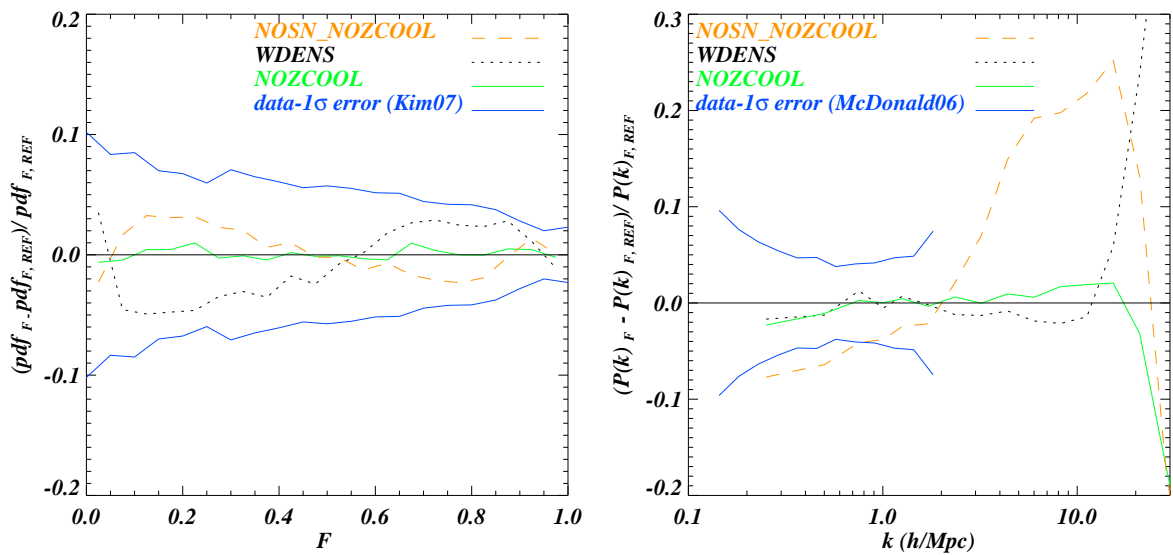}
\end{center}
\caption{Effect of metal cooling, SN feedback and galactic winds in
  terms of flux PDF (left) and flux power (right) for the (25,512)
  simulations at $z=2.25$.  We plot the fractional differences between the flux
  PDF and flux power of the \emph{REF} simulation and the following
  ones: without SN feedback and with primordial cooling (green, solid), with
  strong SN feedback (black, dotted), with SN
  feedback and with primordial cooling (orange, dashed). The blue, solid lines bracket the
  $1\sigma$ statistical observed errors from UVES/VLT at $z=2.07$ PDF
  \citep{tkim} and SDSS flux power at $z=2.2$ \citep{mcdonald06}.}
\label{fig_c}
\end{figure*}

\begin{figure*}
\begin{center}
\includegraphics[width=16cm]{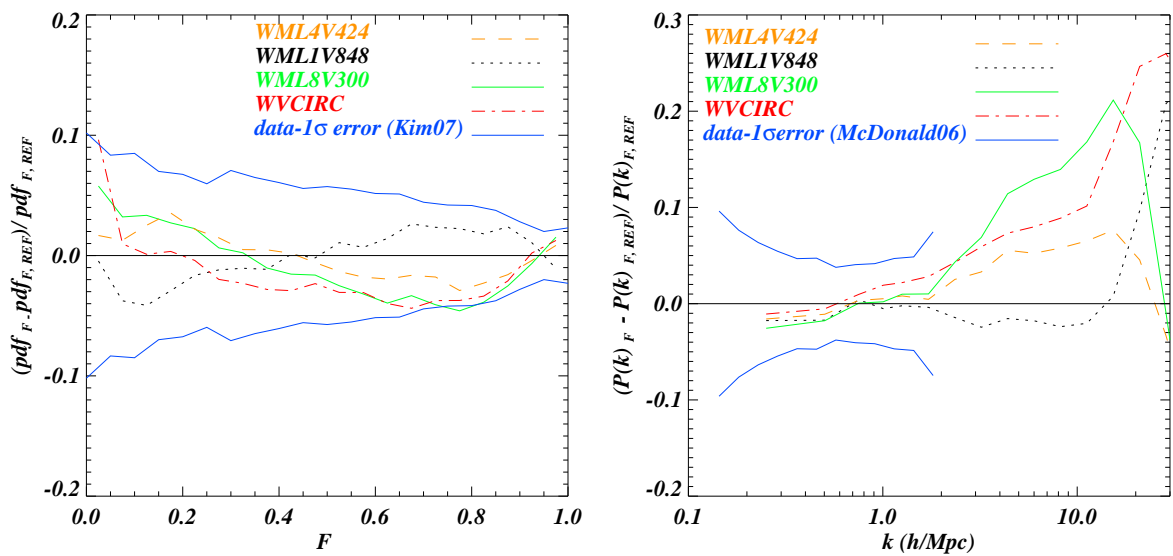}
\end{center}
\caption{Effect of different galactic wind prescriptions in terms of
  flux PDF (left) and flux power (right) for the (25,512) simulations
  at $z=2.25$.  We plot the fractional differences between the flux
  PDF and flux power of the \emph{REF} simulation and the following
  ones: \emph{WML8V300} (green, solid), \emph{WML1V848} (black,
  dotted), \emph{WML4V424} (orange, dashed curve), \emph{WVCIRC}
    (red, dot-dashed curve).  The blue, solid lines bracket the
  $1\sigma$ statistical observed errors from UVES/VLT at $z=2.07$ PDF
  \citep{tkim} and SDSS flux power at $z=2.2$ \citep{mcdonald06}.}
\label{fig_d}
\end{figure*}

\begin{figure*}
\begin{center}
\includegraphics[width=16cm]{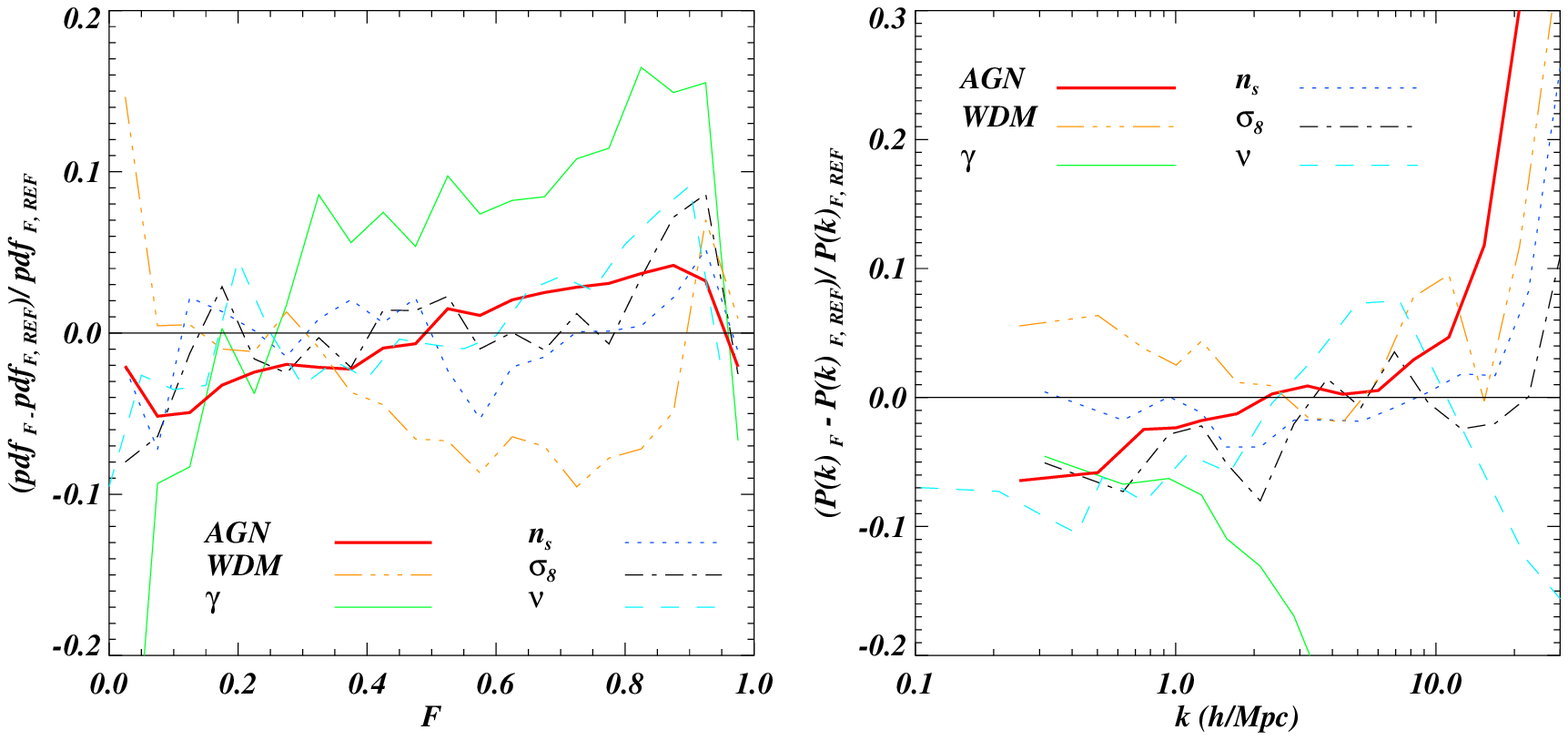}
\end{center}
\caption{Effect of cosmological parameters and thermal state of the
  IGM on the $z=2.25$ flux PDF (left) and flux power (right) compared
  to the AGN effect (fractional differences are shown). We show the
  effect of having: a lower $\sigma_8$ value ($\sigma_8=0.75$)
  compared to the reference case $\sigma_8=0.85$ (black dash-dotted);
  a lower \ns\, value (\ns=0.9) compared to the reference \ns=0.95
  value (blue, dotted); a higher $\gamma$ value ($\gamma=1.6$)
  compared to the reference $\gamma=1.3$ value (green, solid); a
  thermal WDM relic of 1 keV compared to the $\Lambda$CDM case
  (orange, dash-triple-dotted); a neutrino simulation with $\Sigma
  m_{\nu}=0.6$ eV (cyan, dashed) and including AGN feedback (red,
  thick solid). See the text for a more complete description of the
  simulations used.}
\label{figcosmo}
\end{figure*}

In this Section we focus on two different \lya flux statistics that
are widely used for cosmological purposes and for constraining the
physical state of the IGM: the flux PDF and the flux power spectrum.
From each simulation we extract a set of 10000 random
lines-of-sight (LOSs) at $z=2.25$, $z=3$ and $z=4$ and compute the
transmitted flux along each of the LOS in redshift space 
  (unfortunately for the simulations that are not part of OWLS we
  could only sample 1000 LOSs). The synthetic spectra were convolved
with a Gaussian with FWHM $6.6\vel$ and rebinned onto pixels of size
$5\vel$. We generally report our results in terms of ratios or
relative differences with respect to the corresponding reference model
at the same resolution, redshift and box size.

In all the plots shown, we have scaled the optical depth of each
simulation to reproduce the same mean flux level as measured from
observed UVES/VLT high resolution spectra of \citet{tkim}, namely:
${\tau_{\rm eff}}=0.0023 \times (1+z)^{3.65}$ and \teff$=-\ln <F>$
(the value $<F>$ is the mean of the transmitted flux calculated
  using all the pixels of the simulated spectra).  This procedure is
justified because the intensity of the ionising background radiation
is poorly constrained. Note, however, that the original mean fluxes
differ usually by less than one percent between the different
simulations at all redshifts.  In terms of overall scaling factors for
the optical depths, the simulations differ by less than 7\% (the
scaling factors are also very close to one except for the $z=4$
outputs).

We do not consider the impact of resolution and noise on our mock
quasar spectra because these factors do not alter any of our
conclusions with regards to the flux PDF or power spectrum
\citep[e.g.][]{bolton08pdf}.

\subsection{Redshift $z=2.25$}

\subsubsection{Comparison of OWLS models}

In Figures \ref{fig_a} and \ref{fig_b} we show the effect of AGN
feedback on the $z=2.25$ flux PDF (left panel) and flux power (right
panel) for two different box sizes (100,512) and (25,512) represented
by dotted and solid lines, respectively.  Figure \ref{fig_a} shows the
actual PDF and flux power in the two cases, while in Figure
\ref{fig_b} we plot the fractional differences.  The blue lines
bracket the $1\sigma$ error bars obtained from the flux PDF
measurement at $z=2.07$ of \cite{tkim} and the flux power spectrum
measurement at $z=2.2$ of \cite{mcdonald06}.  For the flux PDF the
agreement between the two simulations is at the 2-3\% level. AGN
feedback results in an increase of regions with high transmissivity,
$F>0.5$, at the expense of low-transmissivity regions. This is due to
the fact that, although the simulations share the same $T-\rho$
density relation for the photo-ionised, low-density IGM, the gas is
distributed differently in the $T-\rho$ plane in the two scenarios,
with AGN feedback increasing the amount of low-density gas with
$T>10^5$ K relative to the \emph{REF} simulation.  In this case the
effect induced is comparable to the $1\sigma$ (statistical) error bars
of the measured flux PDF. In terms of flux power, we find a
scale-dependent suppression in the \emph{AGN} case which can reach the
10\% level at the largest scales probed and is $\sim 5$\% at
$k=0.6\,h$/Mpc. We find that AGN feedback is likely to suppress the
flux power at $z\approx 2$ by a scale-dependent amount which is
slightly larger than the $1\sigma$ statistical error bars at
$k=0.2-0.6$ \ihmpc.

It is important to note that the flux power spectra of the two AGN
simulations do not match perfectly due to the large difference in the
numerical resolution of the two runs (a factor 64 in the particle
mass). Our implementation of BH growth is sensitive to the numerical
resolution for BHs with masses similar or smaller than the gas
particle mass \citep{booth09}. While the low-resolution AGN model
reproduces optical and X-ray observations of low-redshift groups of
galaxies \citep{mccarthy10}, the same may not be true for the
higher-resolution run, which was too expensive computationally to run
past $z=2$.

In Figure \ref{fig_c} we show three different effects on \lya flux
statistics: the absence of metal cooling (solid green), the presence
of more efficient SN-driven galactic winds (dotted black) and the
absence of metal cooling and SN feedback (dashed orange). This Figure
refers to $z=2.25$. Turning off metal-line cooling has very little
effect, but if in addition SN feedback is turned off, then there are
substantial differences.  The fact that metal-line cooling
  has little effect in terms of flux properties is expected since the metallicity of the low-density gas that is responsible for the majority of the absorbers is very low \citep{Wiersma11}. Without metal cooling there is a
  decrease of power at very small scale, an expected trend since
  including metal cooling will increase the amount of baryons inside
  the haloes.

Turning off SN feedback suppresses
high-transmissivity ($F=0.6-0.8$) regions, so the effect on the flux
PDF is consistent with that of turning off AGN. Increasing the
strength of SN-driven winds in dense gas (\emph{WDENS}) has a similar
effect on the flux PDF as including AGN feedback. The effects of
feedback processes on the flux power are more confusing: turning on SN
feedback (i.e.\ moving from orange, dashed to green, solid) enhances
the power on large ($k< 2$ \ihmpc) and suppresses it on small scales,
which goes in the opposite direction as the effect of turning on AGN
(see Fig.~\ref{fig_b}). Enhancing the strength of winds in dense gas
has very little effect on the flux power except on relatively small scales.

The fact that the AGN feedback has a different trend in terms of
  flux properties when compared to the SN feedback is due to the fact
  that these two mechanisms have very distinctive behaviours and
  impact the IGM temperature-density relation in a different
  way. In Section \ref{EOS} we will see that removing the contribution
  of gas with $T>10^5$~K to the absorption can result in changes that
  are comparable or larger to those shown by the different feedback
  models. It is therefore likely that the difference between SN and AGN
  feedback is due to thermal and physical differences of the IGM.

In Figure \ref{fig_d} we plot the fractional differences between
\emph{REF} and the three different wind models
\emph{WML1V848,WML4V424,WML8V300} and \emph{WVCIRC} in terms of the flux PDF
(left) and flux power (right) at $z=2.25$.  For the flux PDF the
trends are consistent with the ones shown in Figures~\ref{fig_b} and
\ref{fig_c}: increasing the efficiency of feedback in higher-mass
galaxies (i.e.\ using higher wind velocities) increases the fraction
of high-transmissivity regions at the expense of the
low-transmissivity regions. For the flux power the trend is consistent
with that of Fig.~\ref{fig_c}: more efficient winds decrease the power
on small scales, but enhance it on the larger scales. As noted above,
this is qualitatively different from the effect of AGN feedback, which
suppresses the power on large scales (see Figure \ref{fig_b}).

We note that there is an increasing amount of power at $k>1$ $h/$Mpc
when the mass loading of the winds increases (and the wind velocity
decreases): redistributing the gas outside galactic haloes has the
effect of producing a larger amount of cold gas (detectable in
absorption) at galactic scales that will result in larger amount of
power. In the \emph{WML8V300} and \emph{WML4V424} models more gas is
moved, increasing the power on small scales. In terms of flux PDF
these trends (for kinetic driven winds) are similar to those described
in \citep{bolton08pdf}.

\subsubsection{Other simulations}

In Figure \ref{figcosmo} we compare the effect of AGN feedback to
effects induced by changing some cosmological and astrophysical
parameters.  In this latter case, we use published results that have
been obtained from other sets of simulations. However, even for these
cosmological and astrophysical parameters we compare with a reference
case which is reasonably close to the OWLS \emph{REF} simulation. We
can see that at $z\approx 2$ the effect of AGN feedback on the flux
PDF is comparable to reducing the value of $\sigma_8$ by 0.1 and that
is larger than decreasing \ns\, by 0.05. Changing $\gamma$, the
power-law index of the temperature-density relation
$T=T_0(1+\delta)^{{\gamma}-1}$, from 1.3 to 1.6, or using a WDM matter
power spectrum with thermal $m_{WDM}=1$ keV particles, produces
effects that are larger than that of the feedback and for the case of
WDM they are also of opposite sign.

It is also interesting to compare these models in terms of the flux
power (right panel): the effect of including AGN feedback is quite
scale dependent, at least at scales $k<1 h/$Mpc, while the other
effects could be modelled as an overall suppression or increase of
power.  The magnitude of the AGN feedback effect at the largest scales
is comparable to that induced by a different thermal history and that
produced by a WDM relic. The neutrino induced suppression of the flux
power is of the order of 7-8\% on the largest scales (note, however,
that the neutrino suppression is computed in a model in which the
r.m.s.\ amplitude of the matter power is normalised at the cosmic
microwave background scale).

All the effects not related to the OWLS simulations have been
discussed in previous work: for example the effect of neutrinos, is to
suppress the growth of structure (and the flux power) and thus to
produce a less skewed flux PDF, mimicking a universe with a lower
$\sigma_8$ \citep{vhs10}; the effect of WDM in terms of flux power is
addressed in \citep{seljak06wdm,viel08}, where the fact that it can
increase the flux power at the largest scales is also shown; the
effect of temperature and that of the cosmological parameters with
their redshift evolution is instead addressed in
\citep{mcdonald05,vh06} by using two different hydrodynamic codes.

\begin{figure*}
\begin{center}
\includegraphics[width=14cm]{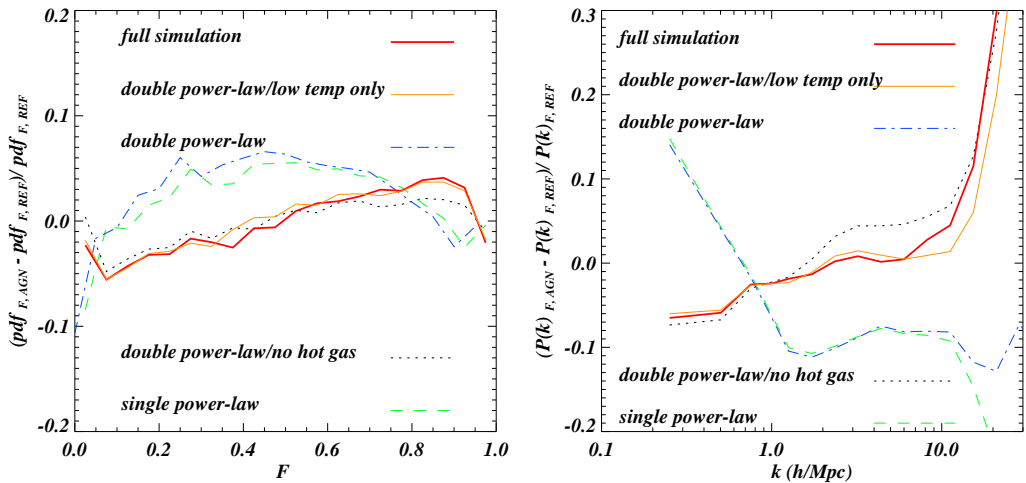}
\end{center}
\caption{Comparison between the \emph{AGN} and \emph{REF} runs at
  $z=2.25$ after imposing an identical $T-\rho$ relation (EOS) on the
  gas in the two models.  The red thick curve refers to the standard
  case in which no temperature rescaling is performed (``full
  simulation''); ``single power-law'' refers to a case in which all
  the gas is placed on the EOS that is appropriate for the
  low-density, photo-ionised IGM (green, dashed curve); ``double
  power-law'' refers to a case in which two different EOS are used for
  the gas below and above an overdensity of $\delta=20$ (blue,
  dot-dashed curve); ``double power-law/low-temp only'' refers to a
  case in which gas with $T>10^5$~K is left unchanged and lower
  temperature gas is placed on a double power-law EOS with separate
  slopes for gas with densities above and below $10^{-4}$~cm$^{-3}$
  ($\delta=20$) (orange, continuous curve); ``double power-law/no hot
  gas'' refers to the same case as ``double power-law'' except that
  the HI fraction of gas with $T>10^5$~K is set to zero (black dotted
  curve). }

\label{fig_i}
\end{figure*}

\subsubsection{Keeping the temperature-density relation fixed}
\label{EOS}
The differences in the flux statistics between simulations using
different feedback prescriptions could be the result of changes in the
distribution of the gas density, changes in the ratio of photo-ionised
and collisionally ionised gas (i.e.\ the fraction of the gas with
$T\gg 10^5$~K), or it could be due to changes in the $T-\rho$ relation
of the photo-ionised gas. As the latter relation can be measured from
the widths of the absorption lines \citep[e.g.][]{Schaye99}, it is
important to check the role of changes in the thermal state of the
diffuse, photo-ionised gas. Below we will follow convention and refer
to the $T-\rho$ relation of the photo-ionised as the ``equation of
state'' (EOS), even though the true equation of state is that of an
ideal gas.

We re-analysed a number of simulations after imposing the same EOS in
a number of different ways. In each case, the EOS is imposed
{\it{a-posteriori}} by changing the temperatures of (a subset) of
particles and the ionization fractions and spectra are then
recalculated. As before, the optical depths are re-scaled so that the
mean flux decrement agrees with the observations of \citet{tkim} for
all models. We consider four different cases:

\begin{itemize}
    \item {\bf Full simulation:} the thermal state of the gas predicted by the simulation is used, i.e.\ no EOS is imposed (red curve, Figure \ref{fig_i}).
    \item {\bf Single power-law:} an EOS of the form
      $T=T_0(1+\delta)^{(\gamma-1)}$ is imposed on all gas at all
      densities, with parameters $T_{0}=10^{4.66}$ K and $\gamma=1.4$
      (green, dashed curve in Fig. \ref{fig_i}).  This EOS fits the $\rho-T$ relation in
      the low-density IGM, but overpredicts the temperatures at high
      density, where radiative cooling is more important than the
      adiabatic cooling due to the Hubble expansion. We include this
      case because a single power-law is a common theoretical
      assumption.

    \item {\bf Double power-law:} At low densities ($\delta<20$) the
      EOS is the same as described above.  At high densities
      ($\delta>20$), the EOS has a slope of $\gamma=0.8$. This double
      power-law EOS provides a good fit to the $T-\rho$ relation of
      the photo-ionised gas in the simulations and also of the gas at
      large densities (blue dot-dashed curve in Fig. \ref{fig_i}).

    \item {\bf Double power-law/low-temp only:} gas at high
      temperatures ($T>10^5$\,K; this gas is mostly collisionally
      ionised) retains its original thermodynamic state.  At low
      temperatures, all gas is placed onto an EOS.  At low densities
      ($\delta<20$) the EOS is the same as described above.  At high
      densities ($\delta>20$), the EOS has a slope of
      $\gamma=0.8$. This double power-law EOS provides a good fit to
      the $T-\rho$ relation of the photo-ionised gas in the
      simulations (orange continuous curve in Fig. \ref{fig_i}).

    \item {\bf Double power-law/no hot gas:} this model is the same as
      the one described directly above, but the density of the hot gas
      ($T>10^5$\,K) is set to zero so that it does not contribute to
      the absorption (black dotted curve in Fig.\ref{fig_i}). This allows us to check the
      effect of absorption by hot gas.

 \end{itemize} 

In Figure \ref{fig_i} we show the fractional differences in the flux
PDF and the flux power between the $z=2.25$ outputs of the \emph{AGN}
and \emph{REF} runs, after both simulations have each of the EOSs
described above imposed. Note that the red, solid curve (``full
simulation'') is identical to the one shown in Figure~\ref{fig_b}.

We can note the following trends. If we impose a ``single power-law''
EOS, then the difference between the \emph{AGN} and \emph{REF} models
is completely different than if we use the predicted temperatures
(``full simulation''). This is because a single power-law strongly
overpredicts the temperature of high-density gas, which now becomes
collisionally ionised and hence mostly invisible in HI. Moreover, the
hot, collisionally ionised low-density gas is now given a low
temperature, which increases its HI fraction. Clearly, imposing a
single power-law EOS on all gas results in large changes to the flux
statistics.

The agreement with the ``full simulation'' is, however, very good if
we use the ``double power-law/low-temp only'' EOS that results in
colder temperature for the high-density gas and which leaves the gas
with $T>10^5~$K unchanged. Imposing this EOS hardly changes the
difference between the \emph{AGN} and \emph{REF} models. The situation
is very different if we impose the double power-law EOS also on gas
that originally had a temperature $T>10^5~$K (dot-dashed blue
curve). In this case the result is more similar to the ``single
power-law'' model than to the ``double power-law/low-temp only''
model.

We can draw two important conclusions from these tests. First, the
differences in the flux statistics between the \emph{AGN} and
\emph{REF} simulations is not due to changes in the EOS of the
photo-ionised gas, because the difference between the two models is
nearly identical if we impose the same, realistic EOS on both
simulations. Second, the difference between models \emph{AGN} and
\emph{REF} is not only caused by changes in the density distribution,
because we obtain completely different results if we impose the double
power-law EOS also on gas with $T>10^5~$K. The fact that AGN feedback
heats gas to temperatures sufficiently high for the gas to become
highly collisionally ionised is very important.

Finally, we note that the gas with $T>10^5~$K is not completely
invisible in HI. We can see this by comparing the cases ``double
power-law; low-temp only'' and ``double power-law/no hot gas''. The
differences are always below the 4\% level in terms of both flux power
and flux PDF and of the order of 1-2\% at the wavenumbers considered
in the SDSS flux power analysis.  Thus, excluding the hot gas at
temperatures above $10^5$ K has a small but non-negligible impact on
the flux statistics.
 
\begin{figure*}
\begin{center}
\includegraphics[width=14cm]{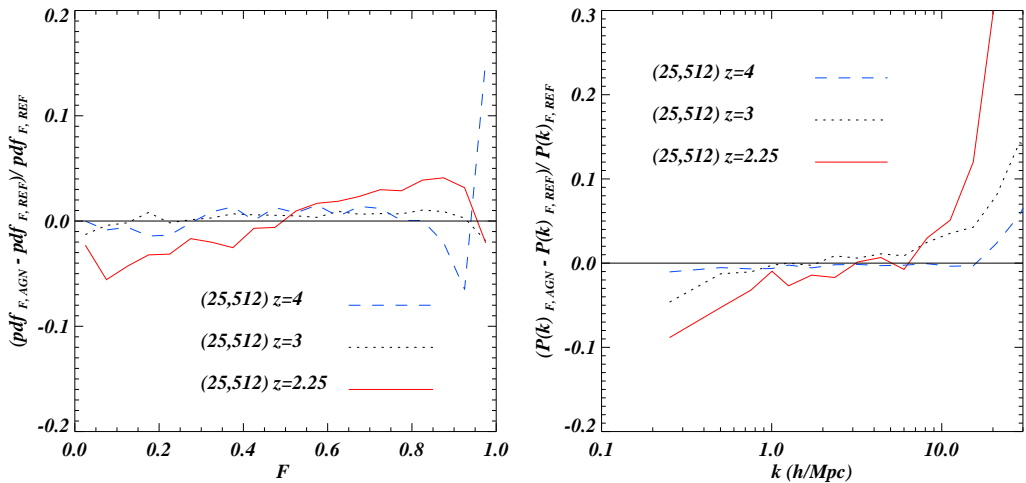}
\end{center}
\caption{Comparison of the effect of AGN feedback on the flux PDF (left) and flux power (right) at redshifts $z=2.25$ (solid, red), $z=3$ (dotted, black), and $z=4$ (dashed, blue). Except for the frequency of very high transmissivity regions, the effect of AGN feedback decreases with redshift.}
\label{fig_e}

\end{figure*}
\begin{figure*}
\begin{center}
\includegraphics[width=14cm]{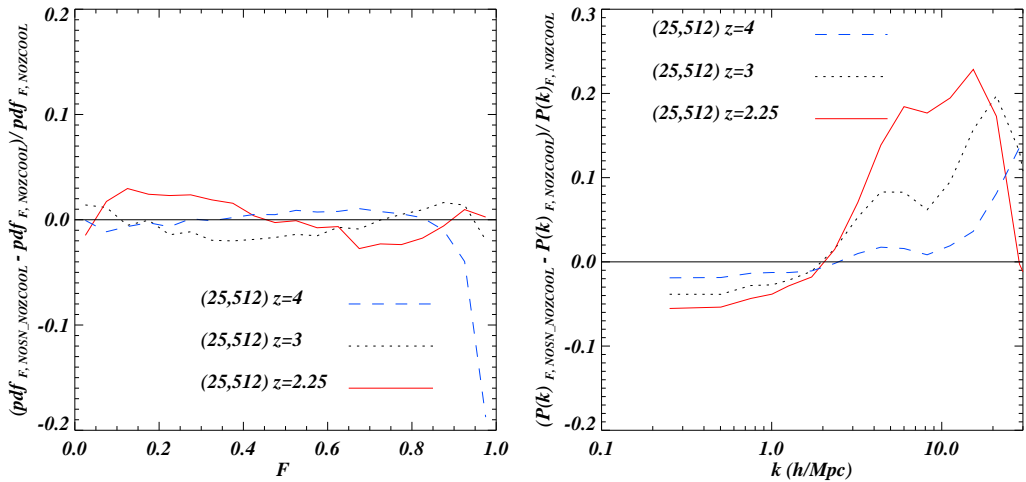}
\end{center}
\caption{Comparison of the effect of SN feedback, using the subgrid
  recipe employed by the \emph{REF} model, on the flux PDF (left) and
  flux power (right) at redshifts $z=2.25$ (solid, red), $z=3$ (dotted,
  black), and $z=4$ (dashed, blue). Except for the frequency of very high
  transmissivity regions, the effect of SN feedback decreases with
  redshift.}
\label{fig_f}
\end{figure*}

\begin{figure*}
\begin{center}
\includegraphics[width=14cm]{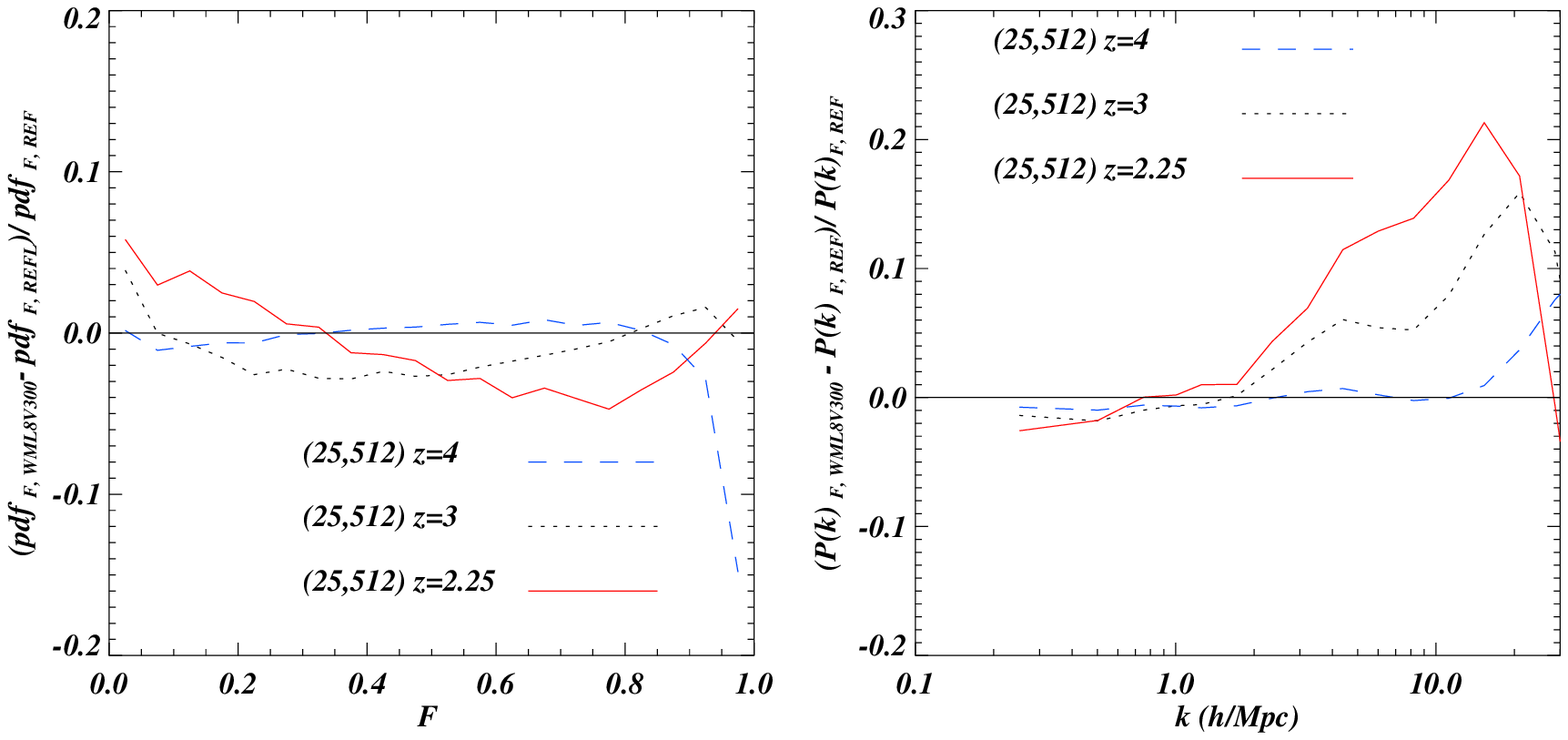}
\end{center}
\caption{Comparison of the difference in the flux PDF (left) and flux
  power (right) predicted by simulation \emph{WML8V300} and \emph{REF}
  at redshifts $z=2.25$ (solid, red), $z=3$ (dotted, black), and $z=4$
  (dashed, blue). Compared with \emph{REF}, the prescription for SN
  feedback employed by model \emph{WML8V300} results in stronger
  outflows in low-mass galaxies. Except for the frequency of very high
  transmissivity regions, the difference between the flux statistics
  predicted by the two simulations decreases with redshift.}
\label{fig_h}
\end{figure*}

\subsection{Evolution}

Up till now we have focused on redshift $z=2.25$, but the flux
statistics have been measured over a substantial redshift range. At
present, PDF measurements span the redshift range $z=2-6$
\citep[e.g.][]{Becker:2006qj} and flux power measurements cover the
range $z=2.1-4.4$ \citep{croft02,vhs,mcdonald06}. The large redshift
range may allow one to lift the degeneracies between cosmological and
astrophysical parameters and results in tight constraints on the
recovered parameters. In this section we will therefore investigate
the evolution of the impact of the physical effects by comparing
predictions for $z=3,4$ with the results for $z=2.25$ that we
already showed in the previous sections.

We caution that the predictions for $z=3$ and especially for $z=4$ are
probably not fully converged with respect to the numerical resolution,
except for the flux power on large scales (see
Appendix~\ref{appendix1resolution}). As we will limit ourselves to
comparing different physical models \emph{at the same resolution},
even predictions that are not fully converged are
informative. However, we must keep in mind that the comparisons may be
affected by the lack of convergence. In particular, the impact of
winds driven by low-mass galaxies may well be underestimated when the
star formation rate in these objects is limited by the finite
resolution of the simulation.

In Figure \ref{fig_e} we compare the \emph{AGN} and the \emph{REF}
simulations at $z=2.25$, 3, and 4. The induced suppression of the flux
power is 3\% smaller at $z=3$ than at $z=2.25$, while at $z=4$ the
models agree since AGN activity, which is restricted to relatively
massive galaxies in our \emph{AGN} simulation, is rare at such high
redshifts. At high redshift there is a substantial difference in the
flux PDF for very high transmissivity regions ($F>0.85$), but this
would be difficult to observe due to continuum fitting errors.

We investigate the redshift dependence of the impact of SN feedback,
for the subgrid recipe employed by the \emph{REF} simulation, by
comparing the \emph{NOZCOOL} and the \emph{NOSN\_NOZCOOL} models at
$z=2.25$, 3, and 4 in Figure~\ref{fig_f}. On large scales, $k<2
h/$Mpc, the effect on the flux power is nearly identical at $z=2.25$
and 3 and at $z=4$ the difference between the two models is less than
2\%. On smaller scales the evolution is stronger. Interestingly, by
$z=4$ the effect of SN feedback only exceeds 2\% for $k > 10
h/$Mpc. The impact of SN feedback on the flux PDF actually increases
with redshift for very high transmissivity regions ($F>0.85$), but as
noted before, continuum fitting errors, which also increase with
redshift, in any case limit the usefulness of such regions.

Figure~\ref{fig_h} shows the evolution of the difference between
simulations \emph{WML8V300} and \emph{REF}. This comparison is
interesting, because for low-mass galaxies the winds in model
\emph{WML8V300} are more efficient than for \emph{REF}. This is
opposite to \emph{AGN}, because AGN feedback only drives strong
outflows in massive galaxies. Nevertheless, by $z=4$ the difference
between \emph{WML8V300} and \emph{REF} is below 2\% for both the flux
PDF and the flux power, except again for the frequency of very high
transmissivity regions and the power on small scales. The \emph{WDENS}
simulation (not shown) has instead very little dependence on redshift
in terms of flux power.  The evolution for
  \emph{WVCIRC} is weaker than for the \emph{WML8V300} model.

We conclude that the three different feedback scenarios have distinct
signatures in the $(k,z)$ parameter space and that the feedback
effects become weaker with increasing redshift, except for the flux
PDF in high transmissivity regions. Although many of the feedback
effects are too weak to be constrained by present data sets, it will
be important to consider them properly when the statistical error bars
are reduced further, as will be the case for ongoing surveys such as
BOSS/SDSS-III.

\section{Conclusions and discussion}
In this paper we have investigated the effect of feedback from galaxy
formation on two widely used \lya flux statistics: the PDF and the
power spectrum of the normalised flux.  We have used state-of-the-art
hydrodynamical simulations from the OWLS project that incorporate most
of the physical processes that may affect the properties of the
IGM. The suite of OWLS runs, of which we have used only a subset, is
particularly well suited for our purposes as the OWLS project was
designed to test the effect of uncertain (subgrid) processes by
varying one parameter at a time in simulations that start from
identical initial conditions.

The impact of feedback on \lya statistics is generically small because
galactic winds tend to escape into the voids, leaving the filaments
responsible for the \lya absorption largely intact
\citep[][]{theunsviel2002}. Indeed, \citet{tepper12} demonstrated this
to be true for the OWLS models by comparing the statistics of Voigt
profile decompositions of synthetic spectra at low redshift. However,
even small differences may be important as the statistical errors on
published data, such as the flux PDF measured from high-resolution
UVES QSO spectra \citep{tkim} and the flux power measured from the
SDSS data \citep{mcdonald06}, are already as small as 5\%. Moreover,
upcoming observational campaigns such as the BOSS/SDSS-III \lya survey
\citep{eisenstein11,slosar11} and the X-Shooter/VLT medium resolution
spectrograph \citep{xshooter} are expected to reduce these error bars
substantially.

We investigated the effect of metal-line cooling, a number of
different implementations of outflows driven by feedback from star
formation, and AGN feedback by comparing different OWLS models to the
OWLS reference simulation (\emph{REF}). For each redshift the spectra
drawn from each simulation were re-scaled to the same, observed mean
flux, thus removing a potential source of differences between
the models (this procedure is justified because the intensity of the
ionising background radiation is poorly constrained). We focused on redshift $z=2.25$ but also studied the
evolution from $z=4$.

Our main conclusions can be summarised as follows:
\begin{itemize}
\item Metal-line cooling has a much smaller effect on the flux statistics than galactic outflows.
\item AGN feedback suppresses the flux power on large scales ($k<1$ \ihmpc). The effect increases in strength with decreasing wave number. At $z=2.25$ it reaches $\sim 10$\% on the largest scales we can measure ($k\sim 0.1$ \ihmpc). The flux PDF is changed at the 5\% level at this redshift. 
\item When using the implementation of \emph{REF}, winds driven by SNe have an effect that is qualitatively opposite to that of AGN feedback. The changes in the PDF and the flux power spectrum caused by SN feedback are slightly smaller than for AGN feedback. 
\item More efficient, but energetically feasible, winds from low-mass galaxies have a strong impact on the flux statistics. At $z=2.25$ the effect is similar, both qualitatively and quantitatively, as that of AGN feedback. At higher redshift the impact of such winds changes and is larger than that of AGN feedback.
\item Overall, the effects induced by feedback from AGN and star formation are of the order of the present 1$\sigma$ statistical error
  bars, comparable to those induced by cosmological and other
  astrophysical parameters.
\item All the feedback mechanisms studied here have a substantially smaller impact on the \lya forest at higher redshift. For winds from low-mass galaxies we can, however, not rule out that the evolution would be weaker for higher resolution simulations. 
\item The feedback effects are not due to changes in the temperature-density relation of the diffuse, photo-ionised gas. Instead, they are due to changes in the density distribution of the gas and to changes in the fraction of hot, collisionally ionised gas at a fixed density.
\end{itemize}

We have shown that the effects of feedback from galaxy formation on
\lya flux statistics are not negligible. In particular, at $z=2.25$
they are comparable to present observed statistical uncertainties. As
upcoming surveys will strongly reduce these statistical errors, it
will become necessary to properly account for these effects in future
analyses of \lya forest data sets.  Although winds driven by feedback
from AGN and star formation are difficult to model from first
principles, progress can be made by using the fact that their effect
on the \lya forest properties  could generically have a different
scale and redshift dependence than changes in the cosmological model.
Further progress in this field can be made by a study at higher
  resolution and by a more quantitative analysis of the effects
  discussed here as a function of redshift and scale.

\section*{Acknowledgments.}
We would like to thank all the members of the OWLS team for their
contributions to this project. The simulations presented here were run
on Stella, the LOFAR BlueGene/L system in Groningen, on the Cosmology
Machine at the Institute for Computational Cosmology in Durham as part
of the Virgo Consortium research programme, and on Darwin in
Cambridge. This work was sponsored by the National Computing
Facilities Foundation (NCF) for the use of supercomputer facilities,
with financial support from the Netherlands Organization for
Scientific Research (NWO), also through a VIDI grant, and from the
Marie Curie Initial Training Network CosmoComp (PITN-
GA-2009-238356). MV is supported by ASI/AAE, INFN/PD51, PRIN-INAF,
PRIN-MIUR and ERC-StG ``cosmoIGM'' grants. This research is also
supported by the European Research Council under the European Union's
Seventh Framework Programme (FP7/2007-2013) / ERC Grant agreement
278594-GasAroundGalaxies and 257670-cosmoIGM. The authors wish to
thank the anonymous referee for the very useful report.

\bibliographystyle{mn2e}
\bibliography{master2.bib}

\appendix

\section{Box-size and resolution effects}
\label{appendix1resolution}

\begin{figure*}
\begin{center}
\includegraphics[width=14cm]{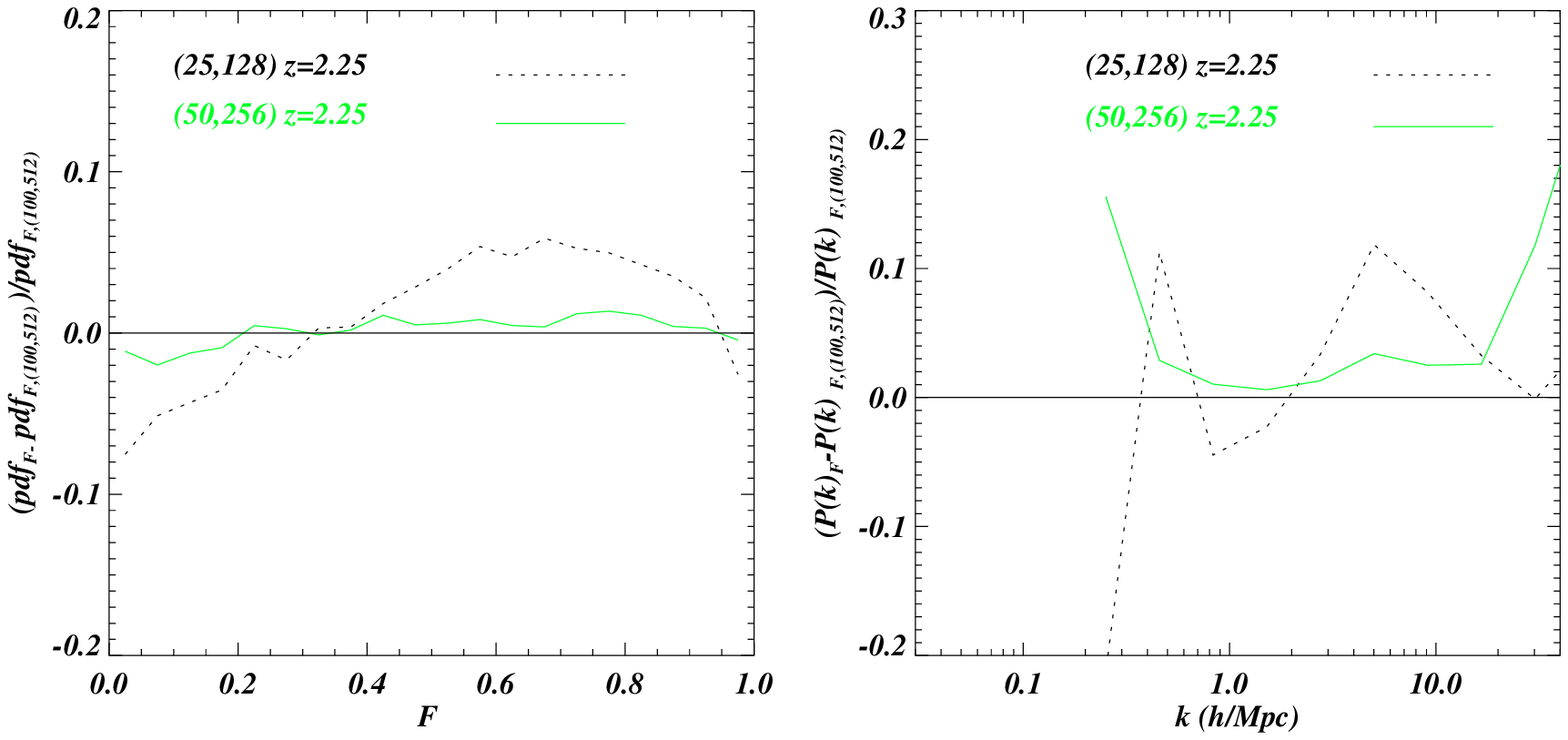}
\end{center}
\caption{Effect of different box size at fixed resolution for the
  \emph{REF} simulations at $z=2.25$. We plot the difference of the
  (25,128) (black, dotted) and (50,256) (green, solid) boxes relative to the (100,512) simulation, respectively (the flux
  power results are only shown for the range of wavenumbers in which
  there is overlap between the different boxes).  }
\label{figbox}
\end{figure*}

\begin{figure*}
\begin{center}
\includegraphics[width=14cm]{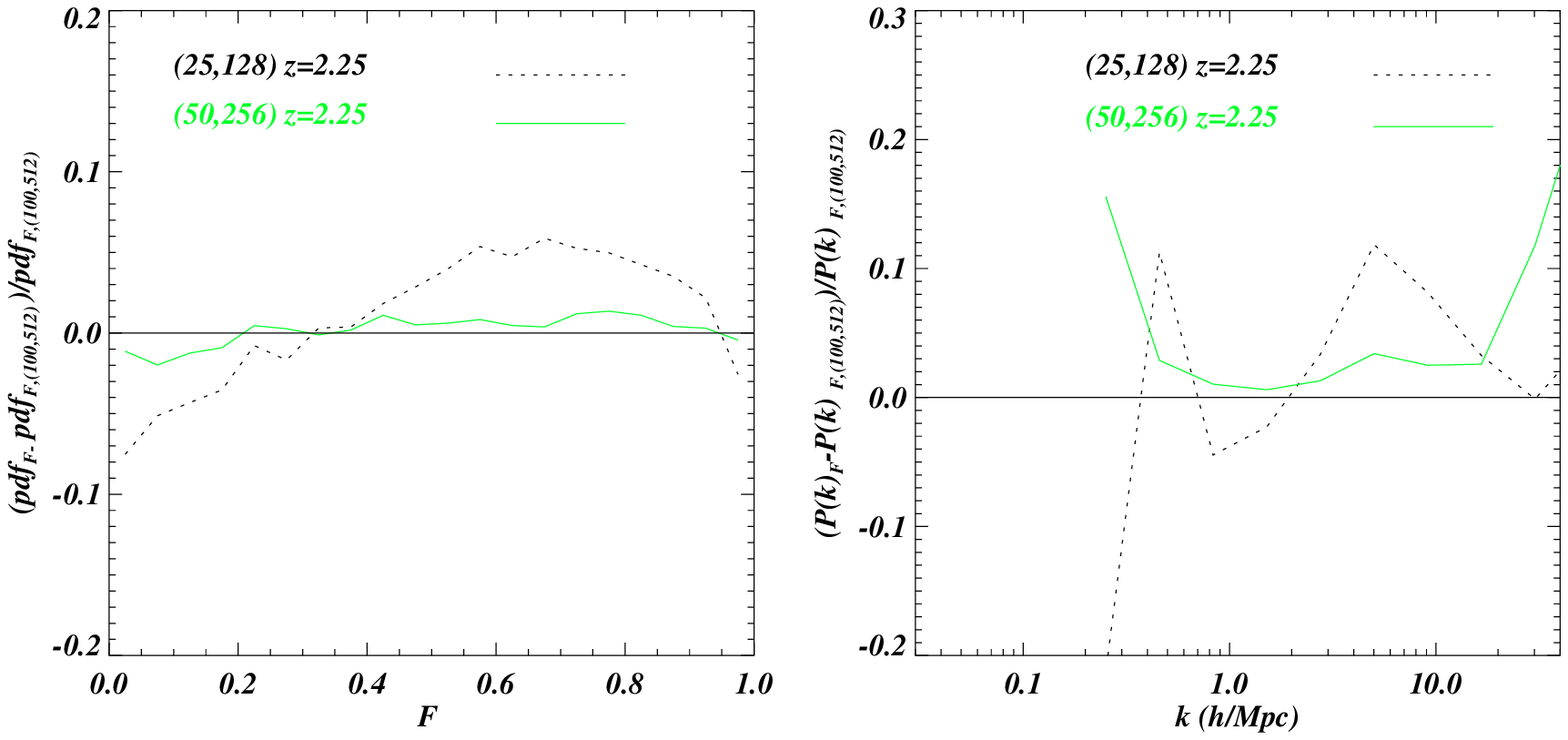}
\end{center}
\caption{Effect of numerical resolution for the reference (\emph{REF})
  simulations at $z=2.25,3,4$. We plot the relative difference of the (25,256) and the (25,512) simulation as the red solid
  ($z=2.25$), dashed orange ($z=3$) and green dotted curves
  ($z=4$). We also show the ratio of the (25,128) and
  (25,512) boxes at $z=2.25$ as a red dot-dashed curve.}
\label{figres}
\end{figure*}

In this appendix, we consider box-size and resolution effects for the
reference (\emph{REF}) runs. In Figure~\ref{figbox}, we quantify the
effects that a different box size has on the flux PDF (left panel) and
on the flux power (right panel) by considering three different boxes
of linear comoving size 100, 50 and 25 \mpch at fixed resolution and at $z=2.25$. From the
Figure we conclude that the 50 and 100 \mpch boxes agree well for $k \ge 1 h/$Mpc. The fact that large scale modes are affected is not
surprising, but this issue is not critical for us since we are interested in the differences of
various physical models relative to the reference model run \emph{in
  the same box size from the same initial conditions}.

Figure \ref{figres} quantifies resolution effects by comparing two
reference runs: the (25,256), represented by the continuous red curve,
and the (25,128), represented by the dot-dashed red curve, to the
reference run at higher resolution (25,512). Note that the mass
(spatial) resolution varies by factors of 8 (2) between each pair of
simulations. At $z=2.25$ the flux PDF is converged to better than 5\%
over the whole flux range and the flux power is also converged at the
3\% level up to $k=3\, h$/Mpc between the (25,256) and (25,512)
runs. At higher redshifts, $z=3,4$ (dashed, orange and dotted, green
curves, respectively) we compare the (25,256) and the (25,512)
simulations and find that the convergence is worse.  We believe that
numerical errors induced by poor resolution are likely to be smaller
than the values quoted above when the (25,512) simulations are
considered (we would require a higher resolution simulation to confirm
this). The fact that we are interested in relative differences between
different models using \emph{the same resolution} will largely
mitigate any resolution effects. An exception may be the
importance of feedback from low-mass galaxies, which we will
underestimate if the star formation rate in low-mass haloes is
limited by the finite resolution. This may be the case at higher
redshifts.
 
\end{document}